\documentclass[prl,superscriptaddress,showpacs,papersize=a4paper,twocolumn]{revtex4-1}
\usepackage[pdftex]{graphicx}
\usepackage{bm}
\usepackage[usenames,dvipsnames]{color}

\newcommand{\bg}{ \begin{gather} }
\newcommand{\eg}{\end{gather}}
\newcommand{\scaling}{6cm}

\begin{document}

\title{Anderson localization and ergodicity on random regular graphs}

\author{K.\,S.~Tikhonov}
\affiliation{L.\,D.~Landau Institute for Theoretical Physics, 142432 Chernogolovka, Russia}
\affiliation{Institut f{\"u}r Nanotechnologie, Karlsruhe Institute of Technology, 76021 Karlsruhe, Germany}

\author{A.\,D.~Mirlin}
\affiliation{Institut f{\"u}r Nanotechnologie, Karlsruhe Institute of Technology, 76021 Karlsruhe, Germany}
\affiliation{Institut f{\"u}r Theorie der Kondenserten Materie, Karlsruhe Institute of Technology, 76128 Karlsruhe, Germany}
\affiliation{L.\,D.~Landau Institute for Theoretical Physics, 142432 Chernogolovka, Russia}
\affiliation{Petersburg Nuclear Physics Institute, 188300 St.\,Petersburg, Russia}

\author{M.\,A.~Skvortsov}
\affiliation{Skolkovo Institute of Science and Technology, 143026 Skolkovo, Russia} 
\affiliation{L.\,D.~Landau Institute for Theoretical Physics, 142432 Chernogolovka, Russia}

\begin{abstract}
A numerical study of Anderson transition on random regular graphs (RRG) with diagonal disorder is performed. The problem can be described as a tight-binding model on a lattice with $N$ sites that is locally a tree with constant connectivity.  In certain sense, the RRG ensemble can be seen as infinite-dimensional ($d\to\infty$) cousin of Anderson model in $d$ dimensions. We focus on the delocalized side of the transition and stress the importance of finite-size effects.
We show that the data can be interpreted in terms of the finite-size crossover from small ($N\ll N_c$) to large ($N\gg N_c$) system, where  $N_c$ is the correlation volume diverging exponentially at the transition. A distinct feature of this crossover is a nonmonotonicity of the spectral and wavefunction statistics, which is related to properties of the critical phase in the studied model and renders the finite-size analysis highly non-trivial. Our results support an analytical prediction that states in the delocalized phase (and at $N\gg N_c$) are ergodic in the sense that their inverse participation ratio scales as $1/N$. 
\end{abstract}

\maketitle
\emph{Introduction.} 
Anderson localization~\cite{anderson58} is a fundamental quantum phenomenon that remains in the focus of current research. A disordered quantum system can be driven (e.g., by increasing disorder) through Anderson transition (AT) between delocalized and localized phase \cite{evers08}. For some class of models (defined on the Bethe lattice, a tree with constant connectivity) the problem of the AT allows for an exact solution, making it possible to establish the transition point and the corresponding critical behavior \cite{abouchacra73,efetov85,zirnbauer86,efetov87,verbaarschot88,mirlin91}.  Recently, the Anderson localization on tree-like graphs has attracted much attention in view of its connections with problems of many-body localization in quantum dots \cite{sivan94a,sivan94,altshuler97,jacquod97,mirlin97,silvestrov97,silvestrov98,DLS01,mejia-monasterio98,leyronas99,weinmann97,berkovits98,leyronas00,rivas02,gornyi16,Kozii16}
and in extended systems with localized single-particle states 
\cite{fleishman80,gornyi05,basko06,ros15,oganesyan07,monthus10,bardarson12,serbyn13,gopalakrishnan14,luitz15,nandkishore15,karrasch15,agarwal15,barlev15,gopalakrishnan15,reichmann15,lerose15,feigelman10,serbyn15,vosk15,ACP15,knap15,ovadyahu,ovadia15,schreiber15,bordia15,Bera15,Geraedts16,bloch16}. 

These developments have motivated Biroli et al.\ \cite{biroli12} to explore the Anderson localization on a tree-like graph (without a boundary) numerically. The authors of Ref.\ \onlinecite{biroli12} have considered a model on a random regular graph (RRG). Since the critical disorder strength $W_c$ does not depend on boundary conditions, the RRG model undergoes the AT at the same point where the corresponding Bethe-lattice model does. For the model considered in Ref.~\onlinecite{biroli12} (connectivity three, hopping set to unity, box distribution of disorder), the transition point was found to be $W_c\simeq{17.5}$ \cite{abouchacra73,monthus09,biroli10}.  The authors  of Ref.~\onlinecite{biroli12} considered the scaling of the level statistics and wavefunction statistics (IPR) with the system size. For conventional disordered systems (in $d<\infty$ dimensions), it is well understood that these quantities have three distinct types of behavior at the localized, critical, and delocalized fixed points \cite{evers08,mirlin00} and have been thus efficiently used to locate the AT \cite{shklovskii93,Hofstetter93,Zharekeshev95,Varga95,Kaneko99,Milde00,Rodriguez09}.
It was observed in Ref.~\onlinecite{biroli12} that the data for matrix sizes $N$ between 512 and 8192 suggest a crossing point at $W_*\simeq{14.5}$. This was interpreted as a possible indication of the intermediate ``non-ergodic delocalized'' phase between $W_T$ and $W_c$, with Poisson level statistics and with the IPR that does not scale as $1/N$. 

Subsequently, the problem of Anderson localization at RRG graphs was considered numerically by De Luca et al.~\cite{deluca14}. These authors focused on the eigenfunction statistics and observed crossing points in singularity spectrum $f(\alpha)$ extracted in a certain way from the distribution of wavefunction amplitudes for systems with $N$ in the range from 2000 to 16000.  On this basis, they conjectured that eigenstates are multifractal (and thus non-ergodic) in the whole delocalized phase, i.e., for all $0<W<W_c$. This would imply, in particular, that the IPR scales in the large-$N$ limit as $P_2\propto{N^{-\mu}}$ with the exponent $\mu(W)$ satisfying $\mu(W)<1$ for all $W<W_c$. 

The possibility of a multifractal delocalized phase in a disordered system is clearly very intriguing. However, the numerical observations of Refs.~\onlinecite{biroli12,deluca14} appear to be in conflict with the analytical predictions of Ref.~\onlinecite{sparse} where the sparse random matrix (SRM) ensemble was introduced and studied analytically. 
It was found that in the delocalized phase and in the limit of large number of sites $N$ (i) the level statistics takes the Wigner-Dyson (WD) form, and (ii)
 the inverse participation ratio (IPR) $P_2=\sum_i|\psi_i|^4$ characterizing fluctuations of an eigenfunction $\psi$ on the infinite cluster (with $\psi_i$ being the wavefunction amplitude on site $i$) scales with $N$ as $P_2\simeq{C/N}$. 
Here the prefactor $C(W)$ depends on the disorder strength $W$, approaching its Gaussian-ensemble value 3 deeply in the metallic phase ($W \to 0$) and diverging as $ \ln C \propto (W_c-W)^{-1/2}$ at the localization transition ($W=W_c$). Numerical results of Refs.~\onlinecite{Sade03,slanina12} for the model on RRG supported the transition from the Poisson to the WD statistics at the AT. 

This has motivated us to perform a detailed analysis of the finite-size scaling of energy-level and wave-function statistics in the Anderson model on RRG, which is the subject of this Letter. 
We have analyzed systems of sizes $N$ from 512 to 65536, with the largest $N$ exceeding those in Refs.~\onlinecite{biroli12,deluca14}.
One of our key observations is a pronounced non-monotonous behavior of observables as functions of $N$ on the delocalized side of the AT. This non-monotonicity, which has a profound origin in the nature of the AT fixed point for tree-like structures (or, equivalently, in the limit $d\to\infty$), makes the finite-size analysis highly non-trivial. Our key conclusion is that the numerical results are fully consistent with the analytical predictions of ergodicity of the delocalized phase (as defined by the WD level statistics and the $1/N$ scaling of IPR in the large-$N$ limit). We expect that our analysis has important implications also for other related problems of Anderson localization, in particular, in the many-body setting.

\emph{Model.} We study non-interacting spinless particles hopping over a RRG with connectivity three in a potential disorder described by the Hamiltonian
\begin{equation}
\label{H}
\mathcal{H}=-\sum_{\left<i, j\right>}\left(c_i^+ c_j + c_j^+ c_i\right)+\sum_{i=1}^N \epsilon_i c_i^+ c_i\,,
\end{equation}
where the sum is over the nearest-neighbour sites of the RRG. The energies $\epsilon_i$ are independent random variables sampled from a uniform distribution on $[-W/2,W/2]$.
Locally looking like a tree, the graph possesses large loops with the typical length of the order of $\log_2N$ \cite{diameter}. This quantity will be referred to as the system size, with the total number of sites $N$ playing the role of volume.
We study the middle of the spectrum ($1/8$ of eigenstates around $E=0$) by exact diagonalization of the Hamiltonian (\ref{H})  and average quantities of interest over disorder realizations (typically over $16000$ to $32$ realizations for $N$ from $512$ to $65536$, respectively).
In what follows, we concentrate on (i) level spacings of neghbouring eigenvalues $\delta_n=E_{n+1}-E_{n}$, and (ii)
wavefunction amplitudes $|\psi^{(n)}_i|^2$.  
As the RRG model differs from that on the Bethe lattice \cite{abouchacra73} only by the presence of very large loops, 
the transition point $W_c$ (defined in the thermodynamic limit $N\to\infty$) is the same in both models, $W_c\simeq{17.5}$.

\emph{Level statistics.} As disorder $W$ passes the transition point, the statistics of the eigenvalues of the Hamiltonian $\mathcal{H}$ qualitatively changes. This transition in the level statistics, which becomes a crossover for a finite system size, has been studied in detail in finite-$d$ models \cite{Hofstetter93,Zharekeshev95,Varga95,Kaneko99,Milde00}. Following Refs.~\onlinecite{oganesyan07,biroli12}, we use as a convenient scaling variable the ensemble-averaged ratio $r=\langle{r_n}\rangle$ of two consecutive spacings,  $r_n = \min(\delta_n,\delta_{n+1})/\max(\delta_n,\delta_{n+1})$, which takes values between $r_{\rm{P}}=0.386$ and $r_{\rm{WD}}=0.530$ realized for the Poisson and the WD Gaussian orthogonal ensemble (GOE) limits, respectively. 

\begin{figure}
  \centering
    \includegraphics[width=0.45\textwidth, height=\scaling]{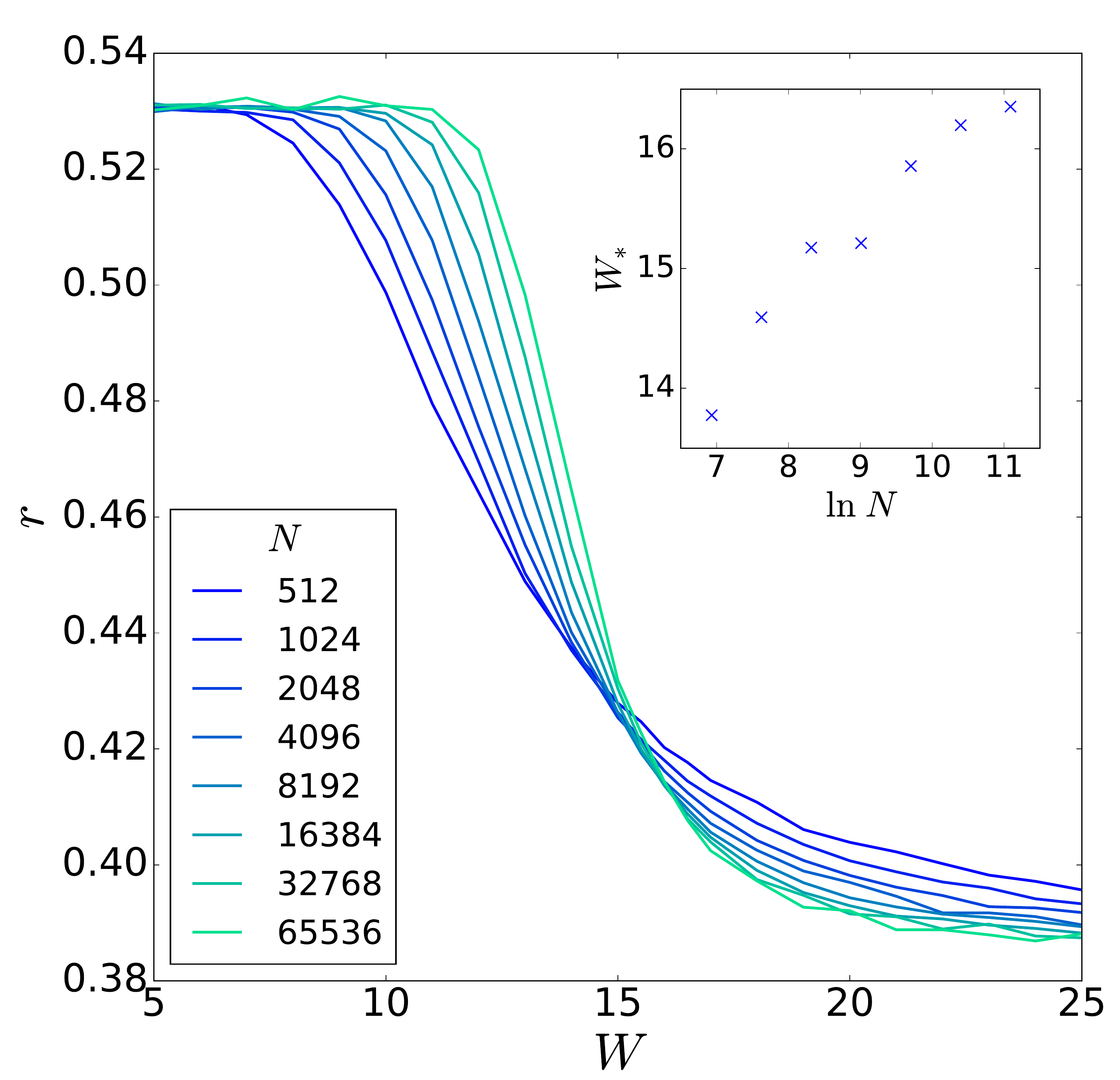}
  \caption{Mean adjacent gap ratio $r$ as a function of disorder $W$ at various $N$ (see legend). Inset: apparent crossing point $W_*$ as a function of the system size, $\ln N$.}
  \label{spsize}
\end{figure}

Our results for dependence of $r$ on $W$ for various $N$ are shown in Fig.~\ref{spsize}. As expected, we observe a crossover from the GOE to the Poisson value that takes place for each $N$ with increasing $W$.  While the crossover becomes sharper for larger $N$, it remains rather broad. This is an indication of the fact that the critical regime [the range of disorder $W$ for which the system size $\log_2{N}$ is larger than or of the order of the correlation length $\xi(W)$] still remains quite broad even for the largest $N=65536$. 

At first glance, the curves in Fig.~\ref{spsize} may seem to show a crossing point somewhere near $W=15$, which is similar to the observation in Ref.~\onlinecite{biroli12}. However, a closer inspection reveals that this apparent crossing point gradually shifts towards larger values of $W$ with increasing $N$.  Specifically, with $N$ increasing from 512 to 65536, the crossing point moves from $W_*\simeq{14}$ to $W_*\simeq{16}$.  This implies that the value of $r$ at the ``moving crossing point'' gradually shifts downwards, i.e., towards the Poisson value.  This shift has a fundamental reason related to the character of the AT critical point on tree-like structures, as we are going to explain.
 
Let us first remind the reader about a character of the AT fixed point in $d$-dimensional systems \cite{evers08,Zharekeshev95,mildenberger02,garcia-garcia07}. In $d= 2+\epsilon$ dimensions (i.e., close to the lower critical dimension $d=2$), the critical point corresponds to weak disorder (or, equivalently, weak coupling, in terms of the effective field theory, the non-linear sigma model), which means that the critical level statistics is close to the WD one and the multifractality is weak. With increasing $d$ the critical point moves towards strong disorder (strong coupling), so that the level statistics approaches the Poisson form and the multifractality takes its strongest possible form in the limit $d\to\infty$. The latter limit corresponds to tree-like models. One of the manifestations of this extreme form of the AT criticality on tree-like structures is the fact that the IPR has a finite limit when the system approaches the critical point from the localized phase \cite{efetov85,zirnbauer86,efetov87,verbaarschot88,mirlin91} (and thus, by continuity, is also finite at criticality). 

Thus, the critical levels statistics in the RRG model is of Poisson form, like in the localized phase. Therefore, contrary to models in finite dimensionality $d$,  there should be no intermediate crossing point for curves $r(W)$ corresponding to different $N$: the crossing point should necessarily drift towards the Poisson limit with increasing $N$. This is exactly what we observe in Fig.~\ref{spsize}. 

\begin{figure}
  \centering
    \includegraphics[width=0.45\textwidth, height=\scaling]{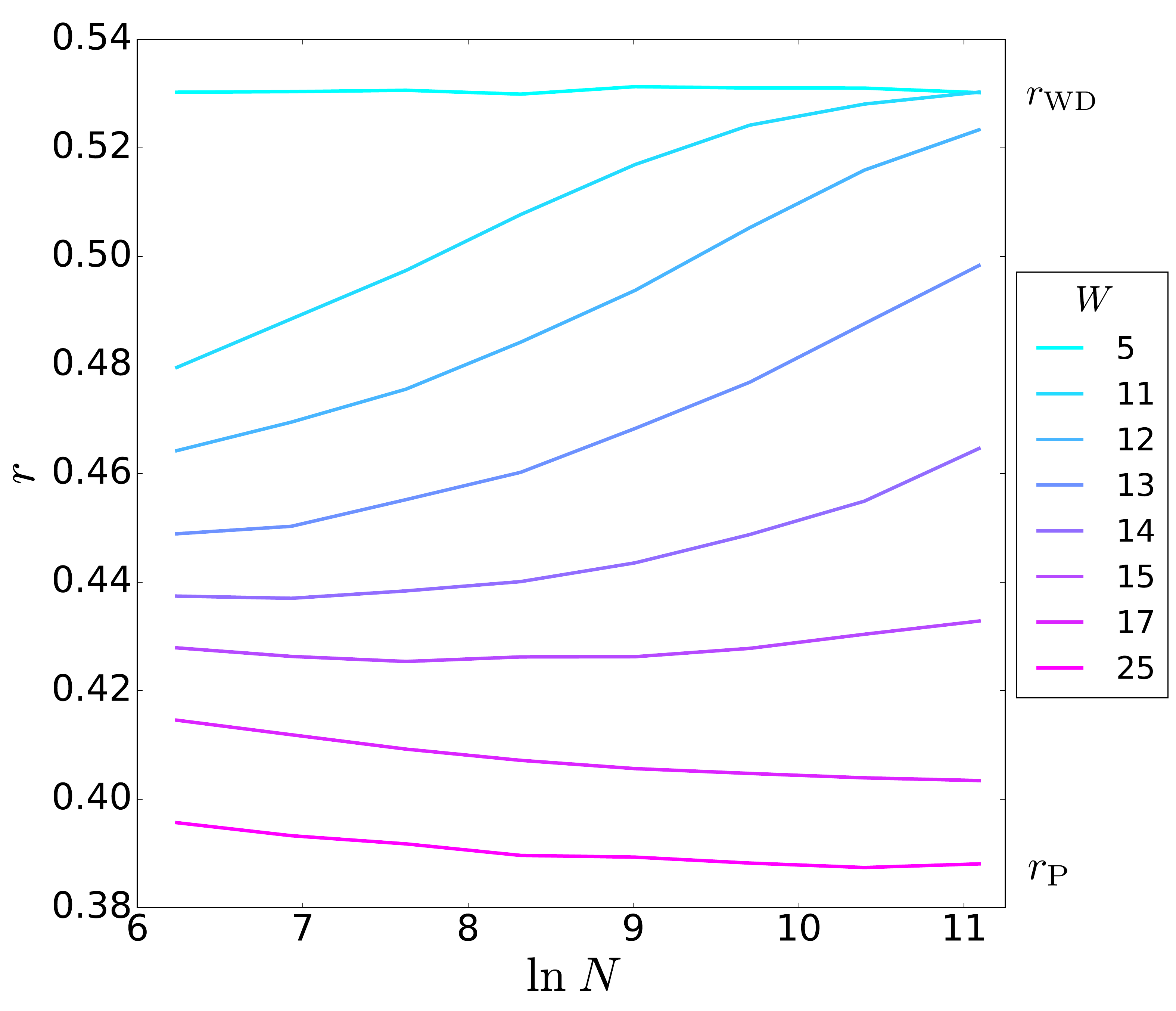}
  \caption{Mean adjacent gap ratio $r$ as function of system size, $\ln N$, at various disorder levels $W$.}
\label{spdis}
\end{figure}

An alternative way to plot the same data is shown in Fig.~\ref{spdis}. Here we show a set of curves  $r(N)$ corresponding to different $W$. The most remarkable feature is a non-monotonous dependence $r(N)$ for curves with moderate disorder on the delocalized side of the transition. The reason for this behavior follows immediately from the above explanation of the shift of the crossing point. Exactly at critical disorder, $W=W_c$, the system develops towards the critical point with increasing $N$, which implies that $r$ decreases, asymptotically approaching its lowest (Poisson) value $r_{\rm P}$. When the system is on the delocalized side ($W<W_c$) but not too far from the transition, it behaves as a critical system as long as its linear size $\log_2 N$ is smaller than the correlation length $\xi(W)$ (diverging in a power-law fashion at $W_c$). Thus, for $N$ smaller than the correlation volume $N_c(W)\sim{2^{\xi(W)}}$, observables develop as at criticality; in particular, $r$ decreases, approaching $r_{\rm P}$. When $N$ reaches $N_c(W)$, the system ``recognizes'' that it is  in the delocalized phase, and $r$ starts increasing towards its large-$N$ limit $r_{\rm WD}$. 
  
\begin{figure}
  \centering
    \includegraphics[width=0.45\textwidth, height=\scaling]{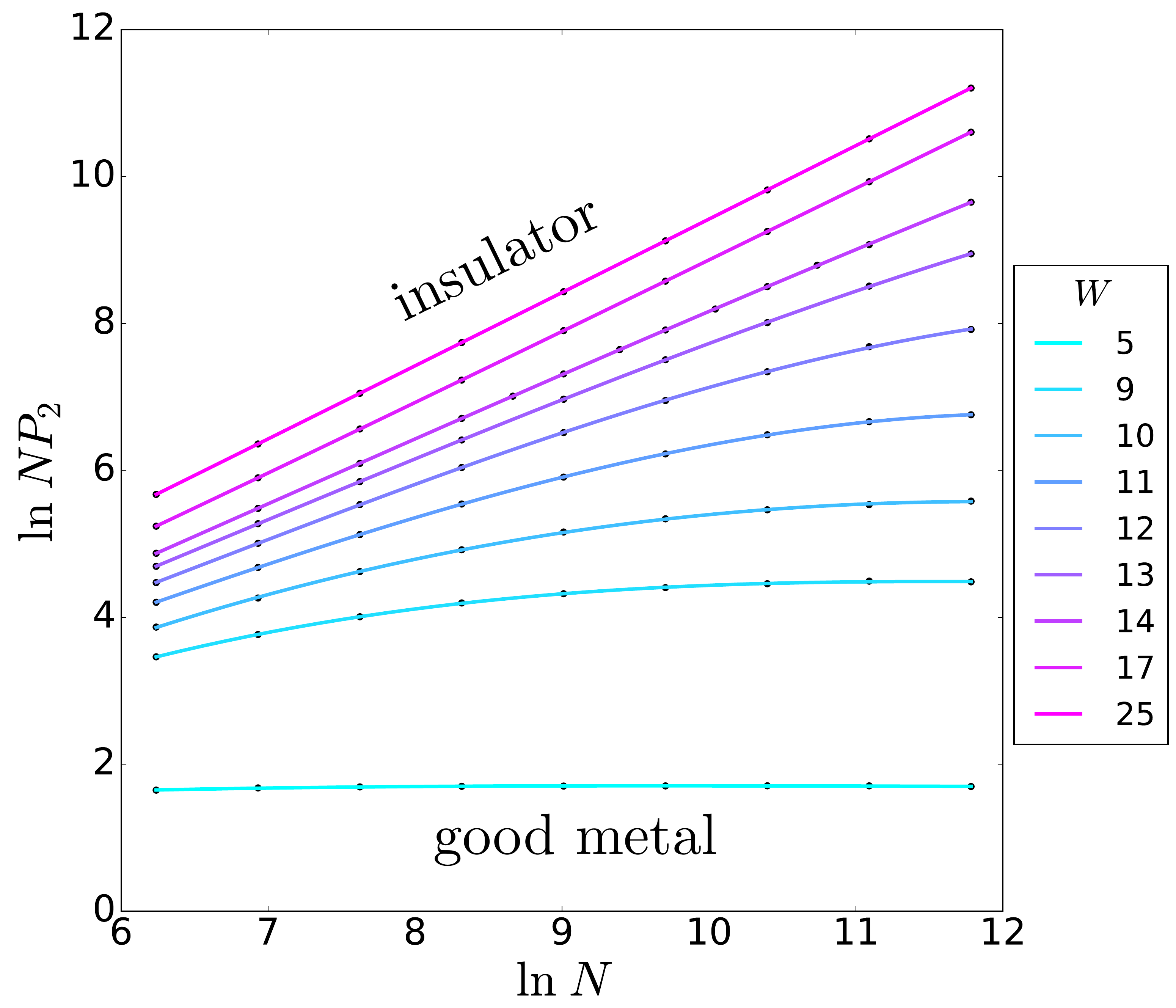}
  \caption{$\ln NP_2$ as a function of the system size at various disorder levels. Dots: simulation, lines: smooth interpolation.}
\label{psisize}
\end{figure}

\emph{Eigenfunction statistics.} Analyzing fluctuations of eigenfunctions, we focus on the (ensemble-averaged) IPR  
$P_2(W,N)=\langle\sum_{i=1}^N\psi^4_i\rangle$.  The system-size dependences of $NP_2(W,N)$ for various strengths of disorder are shown in Fig.~\ref{psisize}. In the localized phase this product increases linearly at large $N$, which is the behavior that the data for $W=25$ clearly show. We consider now the remaining data sets that belong to the delocalized side ($W<W_c\simeq17.5$). 
In the conventional situation expected in the delocalized phase (``ergodicity''), the product $NP_2(W,N)$ saturates at $N\gg1$ at a value $C(W)$ which increases with $W$ approaching $W_c$. This is indeed the behavior that is clearly observed in Fig.~\ref{psisize} when the system is not too close to the transition, $W\lesssim11$. At stronger disorder, $11\lesssim{W}<17.5$, the saturation is not reached
for available system sizes.
The reason is clear from the above 
discussion of the level statistics: the saturation is expected only if $\log_2 N$ significantly exceeds the correlation length $\xi(W)$---the condition that ceases to be fulfilled even for our largest $N$ when the disorder $W$ comes sufficiently close to $W_c$. 

\begin{figure*}
  \centering
    \includegraphics[height=\scaling]{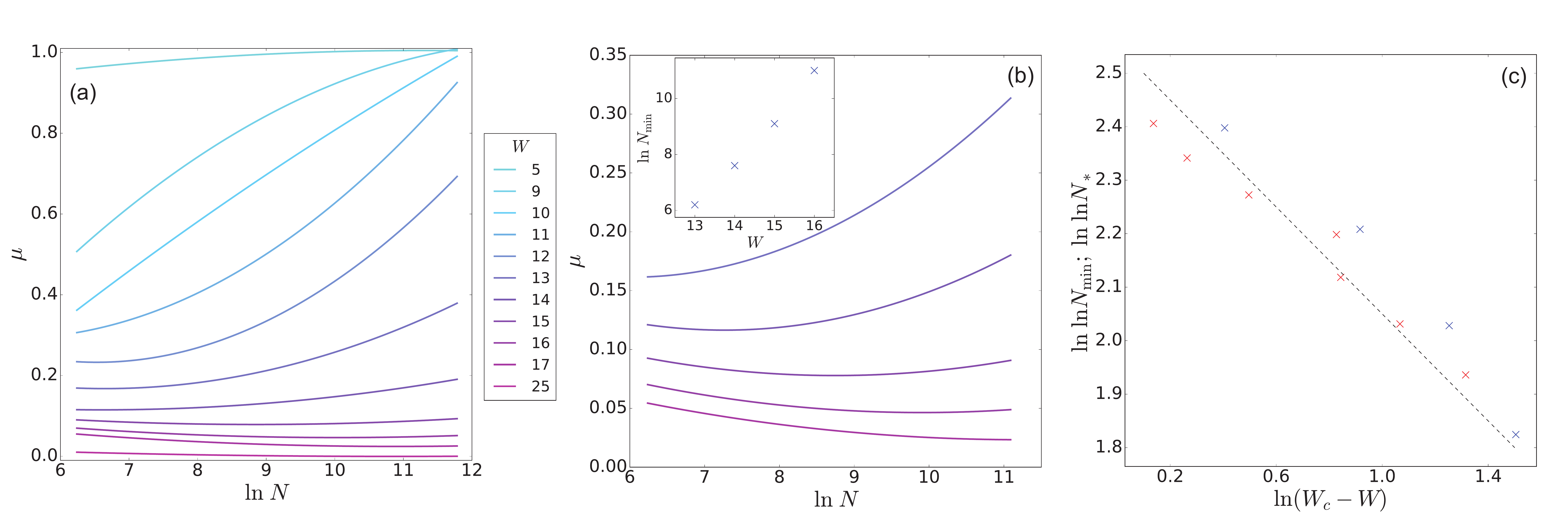}
  \caption{(a) Fractal exponent $\mu$ as a function of the system size at various disorder levels.  (b) Zoom-in for selected disorders $W=13\div17$). Inset: position of the minimum of $\mu(N)$ as a function of disorder. (c) Critical behavior of the correlation length $\xi(W)$ at $W<W_c$ as extracted from position of the crossing point $N_*$ of $r(W)$ curves (red) and minima $N_{\min}$ in $\mu(N)$ (blue). Dashed line corresponds to $\xi(W)\propto(W_c-W)^{-1/2}$.}
\label{alpha-zoomed}
\end{figure*}

It is instructive to replot the data of Fig.~\ref{psisize} by introducing the ``flowing fractal exponent'' $\mu(W, N)=-\partial\ln{P_2}(W,N)/\partial\ln{N}$. The evolution of $\mu(W, N)$  with the system size for various $W$ is shown in Fig.~\ref{alpha-zoomed}a. In this form, the data show a behavior very much analogous to that observed for the level statistics in Fig.~\ref{spdis}. For moderate disorder, $W\lesssim11$, the exponent $\mu$ clearly approaches its ergodic value unity (which corresponds to the saturation in Fig.~\ref{psisize}). For stronger disorder (see Fig.~\ref{alpha-zoomed}b), $11\lesssim W<17.5$, we observe a non-monotonous behavior, the reason for which is exactly the same as has been explained above in connection with Fig.~\ref{psisize}. Specifically, 
$\mu$ first flows towards its value $\mu_c=0$ at the AT critical point. (As pointed out above, this critical point is characterized by a finite value of IPR, as in the localized phase, thus $\mu_c=0$.) When the size $\log_2{N}$ exceeds  the correlation length $\xi(W)$ (see inset in Fig.~\ref{alpha-zoomed}b), the flow turns towards the delocalized fixed point with $\mu = 1$. 

To estimate the critical index $\nu_d$ of the correlation length in the delocalized phase, $\xi(W)\!\propto\!(W_c-W)^{-\nu_d}$, we plot in Fig.~\ref{alpha-zoomed}c the data of insets of Figs. \ref{spsize}, \ref{alpha-zoomed}b versus $W_c-W$ on the logarithmic scale. While the data are not sufficient for an accurate determination of $\nu_d$, they are consistent with $\nu_d = 1/2$, the value suggested by the critical behavior of the IPR \cite{sparse}.

\emph{Summary and discussion.} In this paper, we have studied numerically the level and wave function statistics around the localization transition on RRG (representing a tree-like structure without a boundary), for system sizes $N$ between 512 and $65536$. We have used the mean value  of the ratio of two consecutive level spacings $r$ and the ensemble-averaged IPR $P_2$ (and its logarithmic derivative $\mu$ yielding the ``flowing fractal exponent'')  to characterize these statistics and evaluated their dependencies on $N$ and disorder $W$.
Our main focus was on the behavior on the delocalized side of the transition, $W<W_c\simeq17.5$.  We have found that for moderate disorder 
$W\lesssim11$ our largest system sizes are sufficient to clearly see that the observables reach their conventional (``ergodic'') behavior in the delocalized phase: WD statistics of energy levels and $1/N$ scaling of the IPR. For stronger disorder, $11\lesssim W<W_c$, even our largest system sizes are insufficient to reach the asymptotic large-$N$ behavior. However, we observe in this range of disorder a striking non-monotonous $N$-dependence of observables that strongly supports analytical expectations of ``ergodic'' behavior at large $N$ in the whole delocalized phase.
Specifically, the observables first flow with increasing $N$ towards their critical (AT) values which are of the same character as in the localized phase ($r_c=r_{\rm{P}}$ and $\mu_c=0$). The flow changes its direction when $N$ reaches a value $N_c(W) = 2^{\xi(W)}$ that can be interpreted as a ``correlation volume'': for larger $N$ the observables flow towards their standard values in the delocalized phase, $r_{\rm WD}$ and $\mu = 1$. 
Our results thus corroborate the analytical predictions of Ref.~\onlinecite{sparse} on the WD level statistics and $1/N$ scaling of IPR 
[at $N \gg N_c(W)$] in the whole delocalized phase on tree-like structures. 

Before closing, we make several comments on connections with other works, on further implications of our paper, and on prospective directions for future research.

(i) The fact that extremely large values of $N$ are required to reasonably reach the large-$N$ asymptotics for disorder strengths that are not too close to the critical one (10--20\,\% below $W_c$, i.e., $W=14\div16$) is due to combination of two reasons. First, the condition for the asymptotic behavior is that the linear size $\log_2N$ exceeds the correlation length $\xi(W)$, which requires an exponentially large $N$. Second, even when this condition starts to be fulfilled and the flow---which is initially towards the AT point---turns towards the delocalized fixed point, there is still quite a long way for it, in view of the ``quasi-localized'' character of the AT fixed point on tree-like structures. This relatively large critical window of $W$ at realistic $N$ and the peculiar non-monotonous flow explain the difficulty of the numerical analysis of the problem and a controversy in the recent literature \cite{biroli12,deluca14}. Taking data for $r$ and $\mu$ in a limited range of $N$  may mislead one to a conclusion that the system is ``non-ergodic'' in a certain part of the delocalized phase. 

(ii) For a tree-like model, there is a very essential difference between the system without boundary (like RRG studied in this paper) and a finite piece of tree (with the majority of sites located at the boundary). In the latter case the delocalized states indeed show a (multi-)fractral behavior \cite{monthus11}. We will discuss this difference in more detail separately \cite{tobe}.

(iii) While preparing this work for publication, we learnt about a recent analysis \cite{tarquini16} of the localization transition in the ensemble of L\'evy matrices (LM)---random matrices with entries distributed according to an identical heavy-tailed distribution. While the two problems are quite different,  there is a remarkable similarity in the 
behavior of eigenvalue and eigenfunction statistics near the AT on tree-like structures (RRG studied in the present work)  and in the LM ensemble. 

(iv) Recently, an approach related to that in Ref.~\cite{sparse} was used to calculate finite-$N$ correction to the density of states of RRG ensemble \cite{metz14}. 
Also, Ref.~\onlinecite{metz15} studied the large-energy level statistics. 
A challenging prospective for future work is to study analytically the critical-to-delocalized crossover (with increasing $N$) in the level and eigenfunction statistics for disorder $W$ near $W_c$. 

(v) We expect that some of our results may be relevant also to problems  of the many-body localization transition
in quantum dots \cite{altshuler97,jacquod97,mirlin97,silvestrov97,silvestrov98,DLS01,mejia-monasterio98,leyronas99,weinmann97,berkovits98,leyronas00,rivas02,gornyi16,Kozii16}
and in extended systems
\cite{gornyi05,basko06,ros15,oganesyan07,monthus10,bardarson12,serbyn13,gopalakrishnan14,luitz15,nandkishore15,karrasch15,agarwal15,barlev15,gopalakrishnan15,reichmann15,lerose15,feigelman10,serbyn15,vosk15,ACP15,knap15,ovadyahu,ovadia15,schreiber15,bordia15,Bera15,Geraedts16,bloch16}.
Indeed, numerical studies of this transition do reveal \cite{oganesyan07,luitz15} the drift of the crossing point in the level statistics towards its Poisson limit. As our work demonstrates, such peculiarities make the finite-size scaling analysis a highly challenging task. 

Extrapolating results of our Fig.~\ref{alpha-zoomed}c to $L\equiv\log_2N=100$, we get the uncertainty $(W_c-W)/W_c$ within 1\,\%, suggesting that many-body systems with $\gtrsim 100$ spins (atoms, \ldots) may be sufficient for reaching the transition with high accuracy.
Indeed, recent experiment \cite{bloch16} on a system of $L \simeq 100$ atoms determined quite accurately the position of the many-body localization transition. 

(vi) Recent years have witnessed an impressive progress in mathematical investigations of SRM (Erd\"os-R\'enyi) and RRG ensembles \cite{math}. One thus may hope that analytical results of the type found in Ref.~\onlinecite{sparse} may be cast in a mathematically rigorous form in near future. 

We acknowledge useful discussions with M.V. Feigelman,  A. Knowles, and V.E. Kravtsov.
We thank I.V. Oseledets, L.N. Shchur and L. Yu. Barash for support with the computational power.
The work was supported by the Russian Science Foundation under Grant No.\ 14-42-00044 
and by the EU Network FP7-PEOPLE-2013-IRSES under Grant No. 612624 ``InterNoM''.


\end{document}